\input harvmac


\font\cmss=cmss10 \font\cmsss=cmss10 at 7pt
\def\rlx{\relax\leavevmode}
\def\inbar{\vrule height1.5ex width.4pt depth0pt}
\def\IC{\relax\,\hbox{$\inbar\kern-.3em{\rm C}$}}
\def\IN{\relax{\rm I\kern-.18em N}}
\def\IP{\relax{\rm I\kern-.18em P}}
\def\ZZ{\rlx\leavevmode\ifmmode\mathchoice{\hbox{\cmss Z\kern-.4em Z}}
 {\hbox{\cmss Z\kern-.4em Z}}{\lower.9pt\hbox{\cmsss Z\kern-.36em Z}}
 {\lower1.2pt\hbox{\cmsss Z\kern-.36em Z}}\else{\cmss Z\kern-.4em
 Z}\fi}
\def\IZ{\relax\ifmmode\mathchoice
{\hbox{\cmss Z\kern-.4em Z}}{\hbox{\cmss Z\kern-.4em Z}}
{\lower.9pt\hbox{\cmsss Z\kern-.4em Z}} {\lower1.2pt\hbox{\cmsss
Z\kern-.4em Z}}\else{\cmss Z\kern-.4em Z}\fi}
\font\tenmsx=msam10
\font\sevenmsx=msam7
\font\fivemsx=msam5
\newfam\msxfam
\textfont\msxfam=\tenmsx  \scriptfont\msxfam=\sevenmsx
  \scriptscriptfont\msxfam=\fivemsx
\def\hexnumber@#1{\ifcase#1 0\or1\or2\or3\or4\or5\or6\or7\or8\or9\or%
A\or B\or C\or D\or E\or F\fi }
\edef\msx@{\hexnumber@\msxfam}
\def\llcorner{\,\delimiter"4\msx@78\msx@78 \,}
\def\a{\alpha}
\def\ap{\alpha'}

\def\g{\gamma}
\def\d{\delta}
\def\e{\epsilon}
\def\et{\eta}

\def\l{\lambda}
\def\m{\mu}
\def\n{\nu}
\def\x{\xi}

\def\t{\tau}

\def\p{\phi}

\def\c{\chi}

\def\G{\Gamma}
\def\D{\Delta}

\def\O{\Omega}
\def\om{\omega}
\def\o{\over}
\def\w{\wedge}

\def\frac#1#2{{#1 \over #2}}
\def\pa{\partial}
\def\dg{\dagger}
\def\rd{{ d}}

\def\bx{{\bar \xi}}
\def\bet{{\bar \eta}}
\def\bpa{{\bar \partial}}
\def\bt{{\bar \tau}}
\def\ba{{\bar a}}
\def\bb{{\bar b}}
\def\bc{{\bar c}}
\def\bd{{\bar d}}
\def\bA{{\bar A}}
\def\bB{{\bar B}}
\def\bC{{\bar C}}
\def\aa{\bar 1}
\def\ab{\bar 2}
\def\ac{\bar 3}
\def\dza{dz_1}
\def\dzb{dz_2}
\def\dzc{dz_3}

\def\za{z_1}
\def\zb{z_2}
\def\zc{z_3}

\def\dbza{d{\bar z_1}}
\def\dbzb{d{\bar z_2}}
\def\dbzc{d{\bar z_3}}

\def\bza{{\bar z_1}}
\def\bzb{{\bar z_2}}
\def\bzc{{\bar z_3}}

\def\tJ{{\tilde J}}

\def\CD{{\cal D}}
\def\CI{{\cal I}}


\lref\BB{
K.~Becker and M.~Becker,
``M-theory on eight-manifolds,''
Nucl.\ Phys.\ B {\bf 477}, 155 (1996)
[arXiv:hep-th/9605053];
G.~Curio and A.~Krause,
``Four-flux and warped heterotic M-theory compactifications,''
Nucl.\ Phys.\ B {\bf 602}, 172 (2001)
[arXiv:hep-th/0012152].
}

\lref\BD{K.~Becker and K.~Dasgupta,
``Heterotic strings with torsion,''
JHEP {\bf 0211}, 006 (2002)
[arXiv:hep-th/0209077].
}
\lref\BBDG{
K.~Becker, M.~Becker, K.~Dasgupta and P.~S.~Green,
``Compactifications of heterotic theory on non-K\"ahler complex
manifolds: I,''
JHEP {\bf 0304}, 007 (2003)
[arXiv:hep-th/0301161].
}
\lref\BBDGS{
K.~Becker, M.~Becker, P.~S.~Green, K.~Dasgupta and E.~Sharpe,
``Compactifications of heterotic strings on non-K\"ahler complex
manifolds: II,'' 
Nucl.\ Phys.\ B {\bf 678}, 19 (2004)
[arXiv:hep-th/0310058].
}

\lref\DasRS{
K.~Dasgupta, G.~Rajesh and S.~Sethi,
``M theory, orientifolds and G-flux,''
JHEP {\bf 9908}, 023 (1999)
[arXiv:hep-th/9908088].
}
\lref\Wal{
M.~A.~Walton,
``The heterotic string on the simplest Calabi-Yau manifold and its
orbifold limits,'' 
Phys.\ Rev.\ D {\bf 37}, 377 (1988).
}
\lref\Page{
D.~N.~Page,
``A physical picture of the K3 gravitational instanton,''
Phys.\ Lett.\ B {\bf 80}, 55 (1978).
}
\lref\GauMW{
J.~P.~Gauntlett, D.~Martelli and D.~Waldram,
``Superstrings with intrinsic torsion,''
Phys.\ Rev.\ D {\bf 69}, 086002 (2004)
[arXiv:hep-th/0302158].
}
\lref\Strom{
A.~Strominger, 
``Superstrings with torsion,'' 
Nucl.\ Phys.\ B {\bf 274}, 253 (1986).
}
\lref\Lust{
G.~L.~Cardoso, G.~Curio, G.~Dall'Agata, D.~Lust, P.~Manousselis and
G.~Zoupanos,
``Non-Kaehler string backgrounds and their five torsion classes,''
Nucl.\ Phys.\ B {\bf 652}, 5 (2003)
[arXiv:hep-th/0211118].
}
\lref\GauMPW{
J.~P.~Gauntlett, D.~Martelli, S.~Pakis and D.~Waldram,
``G-structures and wrapped NS5-branes,''
Commun.\ Math.\ Phys.\  {\bf 247}, 421 (2004)
[arXiv:hep-th/0205050].
}
\lref\BergHO{
E.~Bergshoeff, C.~M.~Hull and T.~Ortin,
``Duality in the type II superstring effective action,''
Nucl.\ Phys.\ B {\bf 451}, 547 (1995)
[arXiv:hep-th/9504081].
}
\lref\MesO{
P.~Meessen and T.~Ortin,
``An SL(2,Z) multiplet of nine-dimensional type II supergravity theories,''
Nucl.\ Phys.\ B {\bf 541}, 195 (1999)
[arXiv:hep-th/9806120].
}
\lref\Hassan{
S.~F.~Hassan,
``T-duality, space-time spinors and R-R fields in curved backgrounds,''
Nucl.\ Phys.\ B {\bf 568}, 145 (2000)
[arXiv:hep-th/9907152].
}
\lref\GidKP{
S.~B.~Giddings, S.~Kachru and J.~Polchinski,
``Hierarchies from fluxes in string compactifications,''
Phys.\ Rev.\ D {\bf 66}, 106006 (2002)
[arXiv:hep-th/0105097].
}
\lref\Michel{
M.~L.~Michelsohn, 
``On the existence of special metrics in complex geometry,'' 
Acta\ Math.\ {\bf 149}, 261 (1982).
}
\lref\BT{
K.~Becker and L.-S.~Tseng,
``A note on fluxes in six-dimensional string theory backgrounds,''
Nucl.\ Phys.\ B {\bf 707}, 43 (2005)
[arXiv:hep-th/0410283].
}
\lref\deWSH{
B.~de Wit, D.~J.~Smit and N.~D.~Hari Dass,
``Residual supersymmetry of compactified D = 10 supergravity,''
Nucl.\ Phys.\ B {\bf 283}, 165 (1987).
}
\lref\IvaP{
S.~Ivanov and G.~Papadopoulos,
``A no-go theorem for string warped compactifications,''
Phys.\ Lett.\ B {\bf 497}, 309 (2001)
[arXiv:hep-th/0008232].
}
\lref\KacSTT{
S.~Kachru, M.~B.~Schulz, P.~K.~Tripathy and S.~P.~Trivedi,
``New supersymmetric string compactifications,''
JHEP {\bf 0303}, 061 (2003)
[arXiv:hep-th/0211182].
}
\lref\BecBDP{
K.~Becker, M.~Becker, K.~Dasgupta and S.~Prokushkin,
``Properties of heterotic vacua from superpotentials,''
Nucl.\ Phys.\ B {\bf 666}, 144 (2003)
[arXiv:hep-th/0304001].
}
\lref\Bars{
I.~Bars,
``Compactification of superstrings and torsion,''
Phys.\ Rev.\ D {\bf 33}, 383 (1986);
I.~Bars, D.~Nemeschansky and S.~Yankielowicz,
``Compactified superstrings and torsion,''
Nucl.\ Phys.\ B {\bf 278}, 632 (1986).
}
\lref\Hulla{
C.~M.~Hull,
``Sigma model beta functions and string compactifications,''
Nucl.\ Phys.\ B {\bf 267}, 266 (1986);
C.~M.~Hull and E.~Witten,
``Supersymmetric sigma models and the heterotic string,''
Phys.\ Lett.\ B {\bf 160}, 398 (1985);
}
\lref\Hullb{C.~M.~Hull,
``Compactifications of the heterotic superstring,''
Phys.\ Lett.\ B {\bf 178}, 357 (1986).
}
\lref\KacST{
S.~Kachru, M.~B.~Schulz and S.~Trivedi,
``Moduli stabilization from fluxes in a simple IIB orientifold,''
JHEP {\bf 0310}, 007 (2003)
[arXiv:hep-th/0201028].
}
\lref\GidM{
S.~B.~Giddings and A.~Maharana,
``Dynamics of warped compactifications and the shape of the warped
landscape,'' 
[arXiv:hep-th/0507158].
}
\lref\GukVW{
S.~Gukov, C.~Vafa and E.~Witten,
``CFT's from Calabi-Yau four-folds,''
Nucl.\ Phys.\ B {\bf 584}, 69 (2000)
[Erratum-ibid.\ B {\bf 608}, 477 (2001)]
[arXiv:hep-th/9906070];
T.~R.~Taylor and C.~Vafa,
``RR flux on Calabi-Yau and partial supersymmetry breaking,''
Phys.\ Lett.\ B {\bf 474}, 130 (2000)
[arXiv:hep-th/9912152].
}
\lref\HacL{
M.~Haack and J.~Louis,
``M-theory compactified on Calabi-Yau fourfolds with background flux,''
Phys.\ Lett.\ B {\bf 507}, 296 (2001)
[arXiv:hep-th/0103068].
}

\lref\GraPol{
M.~Grana and J.~Polchinski,
``Supersymmetric three-form flux perturbations on AdS(5),''
Phys.\ Rev.\ D {\bf 63}, 026001 (2001)
[arXiv:hep-th/0009211];
S.~S.~Gubser,
``Supersymmetry and F-theory realization of the deformed conifold with
three-form flux,''
[arXiv:hep-th/0010010].
}
\lref\Yau{
J.~Li and S.-T.~Yau,
``The existence of supersymmetric string theory with torsion,''
[arXiv:hep-th/0411136];
J.~X.~Fu and S.-T.~Yau,
``Existence of supersymmetric Hermitian metrics with torsion on non-Kahler
manifolds,''
[arXiv:hep-th/0509028].
}
\lref\CarCDL{
G.~L.~Cardoso, G.~Curio, G.~Dall'Agata and D.~Lust,
``BPS action and superpotential for heterotic string compactifications with
fluxes,''
JHEP {\bf 0310}, 004 (2003)
[arXiv:hep-th/0306088].
}
\lref\Gukov{
S.~Gukov,
``Solitons, superpotentials and calibrations,''
Nucl.\ Phys.\ B {\bf 574}, 169 (2000)
[arXiv:hep-th/9911011].
}
\lref\GraLW{
M.~Grana, J.~Louis and D.~Waldram,
``Hitchin functionals in N = 2 supergravity,''
[arXiv:hep-th/0505264].
}
\lref\KKLT{
S.~Kachru, R.~Kallosh, A.~Linde and S.~P.~Trivedi,
``De Sitter vacua in string theory,''
Phys.\ Rev.\ D {\bf 68}, 046005 (2003)
[arXiv:hep-th/0301240].
}
\lref\DenDFGK{
F.~Denef, M.~R.~Douglas, B.~Florea, A.~Grassi and S.~Kachru,
`Fixing all moduli in a simple F-theory compactification,''
[arXiv:hep-th/0503124].
}
\lref\IvanP{
S.~Ivanov and G.~Papadopoulos,
``Vanishing theorems and string backgrounds,''
Class.\ Quant.\ Grav.\  {\bf 18}, 1089 (2001)
[arXiv:math.dg/0010038].
}
\lref\TriT{
P.~K.~Tripathy and S.~P.~Trivedi,
``Compactification with flux on K3 and tori,''
JHEP {\bf 0303}, 028 (2003)
[arXiv:hep-th/0301139].
}
\lref\GSW{
M.~B.~Green, J.~H.~Schwarz and E.~Witten,
{\it Superstring Theory: Volume 2, Loop Amplitudes, Anomalies and
Phenomenology,} Cambridge University Press, Cambridge, (1987).  
}
\lref\Joyce{
D.~D.~Joyce, {\it Compact Manifolds with Special Holonomy,} Oxford
University Press, Oxford, (2000).
}
\lref\Chern{
S.~S.~Chern, {\it Complex Manifolds without Potential Theory,} D. Van
Nostrand, Princeton, (1967).
}
\lref\GraMPT{
M.~Grana, R.~Minasian, M.~Petrini and A.~Tomasiello,
``Supersymmetric backgrounds from generalized Calabi-Yau manifolds,''
JHEP {\bf 0408}, 046 (2004)
[arXiv:hep-th/0406137];
M.~Grana, R.~Minasian, M.~Petrini and A.~Tomasiello,
``Generalized structures of N = 1 vacua,''
[arXiv:hep-th/0505212].
}
\lref\HitG{
N.~Hitchin, ``Generalized Calabi-Yau manifolds,''
Quart.\ J.\ Math.\ Oxford Ser. 54, 281 (2003)
[arXiv:math.DG/0209099];
M. Gualtieri, {\it Generalized Complex Geometry,} Oxford University
DPhil thesis, [arXiv:math.DG/0401221].
}
\lref\GraLW{
M.~Grana, J.~Louis and D.~Waldram,
``Hitchin functionals in N = 2 supergravity,''
[arXiv:hep-th/0505264].
}
\lref\work{Work in progress.}
\lref\RohmW{
R.~Rohm and E.~Witten,
``The antisymmetric tensor field in superstring theory,''
Annals Phys.\  {\bf 170}, 454 (1986).
}
\lref\CurKa{
G.~Curio and A.~Krause,
``G-fluxes and non-perturbative stabilisation of heterotic M-theory,''
Nucl.\ Phys.\ B {\bf 643}, 131 (2002)
[arXiv:hep-th/0108220].
}


\Title{\vbox{\baselineskip12pt 
\hbox{hep-th/0509131}
\hbox{MIFP-05-27}
\vskip-.5in}}
{\vbox{\centerline{Heterotic Flux Compactifications and Their
Moduli}}}
\medskip
\centerline{Katrin Becker$\,{}^{1,2,3}$\footnote{$^\dagger$}{{\tt
kbecker@radcliffe.edu}} and $\,$Li-Sheng
Tseng$\,{}^3$\footnote{$^\ddagger$}{{\tt tseng@physics.utah.edu}} }
\bigskip
\centerline{${}^1$ \it George P. and Cynthia W.
Mitchell Institute for Fundamental Physics}
\centerline{\it Texas
A \& M University, College Station, TX 77843-4242,
USA} 
\smallskip
\centerline{${}^2$ \it Radcliffe Institute, Harvard
University, Cambridge, MA 02138, USA}
\smallskip
\centerline{${}^3$ \it Department of Physics,
University of Utah} \centerline{\it Salt Lake City, UT 84112-0830,
USA}

\baselineskip18pt
\medskip\bigskip\medskip\bigskip
\baselineskip16pt

\centerline{\bf Abstract}
\vskip 0.15in

We study supersymmetric compactification to four dimensions with
non-zero H-flux in heterotic string theory.  The background
metric is generically conformally balanced and can be conformally K\"ahler if
the primitive part of the H-flux vanishes.  Analyzing the linearized
variational equations, we write down necessary conditions for the
existence of moduli associated with the metric.  In a heterotic model
that is dual to a IIB compactification on an orientifold, we find the
metric moduli in a fixed H-flux background via duality and check that
they satisfy the required conditions.  We also discuss expressing the
conditions for moduli in a fixed flux background using twisted differential
operators.


\Date{September, 2005}


\newsec{Introduction}

One of the main appealing features of flux compactifications in string theory
is its ability to stabilize moduli fields.  From the perspective of  
the low energy effective action, turning on fluxes generates a
classical potential
\refs{\GukVW,\Gukov,\HacL,\GidKP,\BecBDP,\CarCDL} which
typically lifts many of the moduli.  Any remaining moduli may be further
stabilized by non-perturbative effects \refs{\CurKa,\KKLT}.  (See
\DenDFGK\ for an
explicit example.)  Such an approach assumes that the back-reaction of
the non-zero fluxes on the background geometry is mild and that the
geometry in the presence of fluxes can be continuously deformed
from one without flux.  This is certainly the case in well-studied M-theory 
\BB\ and type IIB \GraPol\ flux models where the background metric is
modified simply by an additional warp factor when the fluxes are turned on.

However, in supersymmetric heterotic compactifications with non-zero
H-flux, the back-reaction on the geometry can be drastic.  Typically,
the non-zero flux background geometry is non-K\"ahler, and more
significantly, they are also topologically different from the
zero-flux Calabi-Yau manifold \refs{\BBDG, \BecBDP}.  Such background
geometries necessarily can not be continuously deformed from a Calabi-Yau
geometry \refs{\BD, \CarCDL}.  The excitations of the low-energy
effective action, in particular the moduli fields, are no longer justifiably
those present prior to the flux being turned on.  An
important issue then is to identify the moduli fields directly from
the new flux geometry. 

In this paper, we begin the study of the moduli fields in
supersymmetric heterotic compactifications on six-dimensional
non-K\"ahler geometries.  Our analysis will be confined within the
framework of supergravity.  The study of moduli is then simply 
a linearized variational problem.  The background fields satisfies
certain constraining differential equations (e.g. Killing spinor
equations or equations of motion) and the moduli corresponds to
infinitesimal variation of the fields such that the constraining
equations remain satisfied.  Unfortunately, the differential
equations though linear in variations typically couple the various
variations in a complicated manner.  Thus, solving the variational
problem in general can be rather challenging.

In our study, we are aided by the fact that the currently known explicit
models of heterotic flux backgrounds are dual to orientifold flux
models in type IIB string theory \refs{\DasRS, \BD}.  Disregarding
non-perturbative effects, the IIB orientifold models contain unlifted
K\"ahler moduli in the presence of non-zero fixed $G_3$ fluxes.  These IIB
moduli can be mapped to give moduli in the heterotic flux model.  A main
motivation of ours is to understand these moduli solely within the
context of the heterotic theory.  The moduli are those associated with
the metric and with the H-flux held fixed in the variation.  Hence, our
analysis of the heterotic moduli will concern mainly with the metric
moduli and will also emphasize those in a fixed H-flux background.  

We begin in section 2 by giving an overview of the possible
six-dimensional geometries that preserve $N=1$ supersymmetry in four
dimensions.  In general, with non-zero H-flux, the metric is required
to be conformally balanced.  However, if the primitive part of the H-flux
vanishes, then the metric is conformally K\"ahler.  We then proceed to perform
a linearized variation on the supersymmetry constraints for the background
fields.  We focus on the moduli associated with the metric. 
A metric variation will generically also require a simultaneous
variation of the dilaton field.   We will write down necessary
conditions which the moduli associated with the metric must satisfy.
In section 3, we study the moduli 
of the heterotic flux models which are dual to models of IIB orientifolds with
fluxes.  We first write down the unlifted K\"ahler moduli in the IIB
theory and then map them via supergravity duality rules to the
heterotic theory.  We find that these moduli satisfy the required
conditions.  We conclude in section 4 commenting on some open
questions and discussing the two conditions that the metric moduli in
a fixed flux background must satisfied.  Interestingly, the conditions
may be re-expressed in terms of a twisted differential operator
similar to those introduced in \refs{\HitG, \GraMPT}.

\newsec{Moduli of heterotic compactifications with H-flux}

We first review the constraints on supersymmetric flux
compactification on $M^{3,1}\times K$, a four-dimensional Minkowski
spacetime times a six-dimensional internal spin manifold
\refs{\Strom,\Lust,\GauMW} (see also \refs{\Bars,\Hulla,\Hullb}).  The
modern classification of supersymmetric backgrounds is given in terms
of intrinsic torsion classes \refs{\Lust, \GauMW}.  However, for the
heterotic theory, the intrinsic torsion is determined solely by the
H-flux.  Thus, instead of linking the background geometries with intrinsic
torsion classes, we emphasize the required geometry for each type of
H-flux present.   We take note of an interesting subset of solutions 
with conformally K\"ahler metrics on Calabi-Yau manifolds.  We then
proceed to study the moduli in backgrounds with non-zero H-flux.

\subsec{Constraints on Heterotic Solutions}

The background fields of a supersymmetric compactification in
heterotic theory must obey 
\eqn\heta{\d \psi_M = \nabla_M \e + {1\o 8} H_{MNP}\,\G^{NP}\, \e=0~,}
\eqn\hetb{\d \l  = \G^M\pa_M\phi\, \e + {1\o 12} H_{MNP}\,\G^{MNP}\,\e=0~,}
\eqn\hetc{\d \c = \G^{MN}F_{MN} \e =0~, }
for some positive chirality 10d Majorana-Weyl spinor $\e$.\foot{We mostly follow the conventions of \GauMW.  The gauge field strength $F_{MN}$ is however here taken to be anti-hermitian.}  In
addition, the H-field obeys the modified Bianchi identity
\eqn\hetd{dH = {\ap \o 4} ({\rm tr}\,R^{(-)}\w R^{(-)} - {\rm tr}\,
F \w F)~.}
As for quantization of the H-flux, with $d H\neq 0$, $H$ is in general
not an element of the integer de Rham cohomology class
$H^3(K, \ZZ)$.  However, if there exist three-cycles in regions of the
manifold where $dH=0$, then there is a Dirac quantization
\eqn\dirac{{1\o 4\pi^2 \a'}\int_\G\, H \in \ZZ~.}   
where $\G$ is any three-cycle in those regions.  Outside these region,
the quantization condition is modified by the inclusion of a ``defect''
term $\d= (1/ 16\pi^2)\int_\G\,[\Omega_3(A) - \Omega_3(\om^{(-)})]$ where
$\Omega_3$ is the Chern-Simons three-form \RohmW. 
In general, this defect term does depend on the particular three-cycle $\G$.

We note that the four equations \heta-\hetd\ together are sufficient
conditions for a background field configuration to satisfy the
equations of motion of the 10d supergravity theory \GauMPW.   We shall
take the spinor ansatz to be 
\eqn\sdecomp{\e = \x \otimes \et\, + \,\bx \otimes \bet,}
where $\x$ and $\et$ are 4d and 6d Weyl spinors, respectively, and
$\bx=(B^{(4)}\x)^*$ and $\bet=(B^{(6)}\et)^*$ are the complex
conjugated spinors.\foot{$B^{(4)}$ and $B^{(6)}$ are the complex
conjugation matrices that complex conjugate gamma matrices in four and
six dimensions, respectively.}   $\et$ will be taken to have positive
chirality
and unit normalized, $\et^\dg\,\et =1 $.  As a supersymmetry
transformation parameter, $\et$ must be everywhere non-vanishing on
the internal manifold $K$.  The presence of a single such spinor 
implies that there exists on $K$ a connection with $SU(3)$ holonomy. 
For $H\neq 0$, the $SU(3)$ holonomy connection is no longer the
Levi-Civita connection.  It has non-zero torsion and is precisely
the torsional connection, ${\om^{(+)}}_{M\;B}^{\ \ A} =
{\om}_{M\;B}^{\ \ A}+{1\o 2} {H}_{M\;B}^{\ \ A}$, as given in \heta, 
for which $\et$ is covariantly constant.\foot{A manifold with an $SU(3)$
holonomy connection has an $SU(3)$ structure group.  Also, in the
mathematics literature, the torsional connection is called the
Bismut connection.}  We also point out that the curvature tensor in
\hetd\ is defined with
respect to the ``minus'' connection, $\om^{(-)}= \om - {1\o 2} H$.
The justification comes from the worldsheet non-linear sigma
model where the $\om^{(-)}$ connection is required
to preserve both worldsheet conformal invariance and spacetime
supersymmetry \Hullb.

We will focus on the implications of the first two constraint equations,
\heta\ and \hetb.  Together, they imply that $K$ is a complex manifold
with the complex structure
\eqn\jform{{J_m}^n = - i\, \et^\dg\, {\g_m}^n \,\et~,}
which satisfies ${J_m}^k{J_k}^n=-{\d_m}^n$.  The metric is hermitian with
respect to the complex structure,
i.e. $g_{mn}=J_m{}^rJ_n{}^sg_{rs}\,$, and is used to define the
associated real hermitian two-form $J_{mn}={J_m}^kg_{kn}\,$.\foot{To avoid
any confusion, we refer to $J_{mn}$ as the hermitian form, following Joyce
\Joyce.  As in \Chern, it is also called the K\"ahler form even when
the K\"ahler condition, $\rd J=0\,$, is not satisfied.} 
As for the three-form flux, $H$, the two constraints imply the
following equations
\eqn\Hcondaa{H_{mnp}= 3{J_{[m}}^r\nabla_{|r|}J_{np]}~,}
\eqn\Hcondb{H_{mnp}\,J^{np} = -2\, \nabla_p {J_m}^p = -4 \,
{J_m}^p\,\pa_p\phi~.}

We will work in complex coordinates, $z^a$ and ${\bar z}^\ba$, such that
the complex structure takes the standard form,  ${J_a}^b=i{\d_a}^b$,
${J_\ba}^{\bb}=-i{\d_\ba}^{\bb}$, and with other components zero.  The
hermitian form, $J_{mn}$, is now easily seen to be a $(1,1)$-form
\eqn\hmet{J_{a\bb} = -J_{\bb a}= i\, g_{a\bb}~,}
and the constraint on $H$ \Hcondaa\ simplifies to
\eqn\Hconda{H=i(\bpa-\pa)J}
where $J$ denotes the hermitian form and for
example, $H_{ab\bc}=-i(\pa_a J_{b\bc} -\pa_b J_{a\bc})$.  From
\Hconda, we see that the real three-form H-flux must 
be a $(2,1)$- and $(1,2)$-form with respect to the complex structure.
Furthermore, with 
$H^{(2,1)}=-i [\rd J]^{(2,1)}$, it is clear that a non-zero H-flux
forces the internal manifold $K$ to be non-K\"ahler.  As for \Hcondb, it
describes the primitivity of $H$.  We see that $H$ is only primitive
(i.e.~$H_{mnp}J^{np}=0$)  
if and only if the dilaton is a constant.  But as noted in
\refs{\IvanP, \GauMW}, the equations 
of motion of $N=1$ 10d supergravity can only be satisfied for a
constant dilaton if also $H=0$.  Thus, for a non-zero H-flux
background, $\phi$ must be non-constant and $H$ is necessarily
non-primitive.\foot{The requirement that $H$ is non-primitive is a
supergravity condition, valid at lowest order in $\alpha'$.  Including
$\alpha'$ corrections may possibly loosen this requirement.  An
example of a constant dilaton background that satisfy the supersymmetry
constraints \heta-\hetc\ but not the equation of motion is the Iwasawa
manifold solution given in \Lust.  As noted in \GauMW, the solution
does not satisfy the modified Bianchi identity \hetd.}
It is useful to decompose $H$ as a sum of its primitive $H^P$ and
non-primitive $H^{NP}$ parts as follows
\eqn\Hpnp{\eqalign{H_{mnp}&= H^P_{mnp} + H^{NP}_{mnp} \cr
&=\left(H_{mnp} - {3\o 4}\, J_{[mn}H_{p]rs}J^{rs}\right) + {3\o 4}\,
J_{[mn}H_{p]rs}J^{rs}~. }}
where $H^{P}_{mnp}\, J^{np} =0$.  Applying \Hcondb, we have that
\eqn\Hnp{H^{NP}= i(\bpa-\pa)\phi\w J = -*(\rd \phi \w J)}
where the Hodge star, $*$, is defined with respect to the volume form ${1\o
3!} J\w J\w J$.  And with \Hconda, we find that $\rd J$ also has a non-zero
non-primitive part given by
\eqn\dJnp{\rd J^{NP}= \rd\phi \w J~.}  

Finally, the two constraints also imply the existence of a holomorphic
three-form.  Consider the three-form fermion bilinear
\eqn\Oform{\O_{mnp} =  \bet^\dg \g_{mnp}\, \et~,} 
which is a $(3,0)$-form with respect to the complex structure of \jform.
Utilizing the two constraints, the anti-holomorphic derivative acting
on $\O$ gives 
\eqn\Ocond{\pa_\ba(\O_{abc}) = 2\, \pa_\ba\phi\ \O_{abc}~,}
and therefore, we have that $w_{abc}= e^{-2\phi}\O_{abc}$ is a holomorphic
three-form.  The existence of such a three-form is due to the
internal manifold $K$ having $SU(3)$ holonomy with respect to the
torsional connection.  The $SU(3)$ holonomy also implies that $K$ has
zero first Chern class, i.e. $c_1(K)=0$.

The above constraints on the fields, \Hconda, \Hcondb, and \Ocond, can be
expressed simply in terms of exterior derivatives and the Hodge star
operator as follows \GauMW
\eqn\heqa{ H= * \,e^{2\phi}\rd(e^{-2\phi} J) ~,}
\eqn\heqb{ \rd(e^{-2\phi}\,* J)=\rd({e^{-2\phi}\o 2} \,J\w J)=0~,}
\eqn\heqc{ \rd(e^{-2\phi}\O)=0~.}
The equivalence of \heqa\ with \Hconda\ can be verified using \Hnp\
and that in six dimensions, a primitive $(p,q)$-form satisfies the
self-duality relation $*A^{(p,q)}=i(-1)^q A^{(p,q)}$ for
$p+q=3$.\foot{In our convention, $(*A)_{mnp}={1\o 3!}\e_{mnp}{}^{qrs}A_{qrs}$ with the Levi-Civita tensor $\e_{mnpqrs}$ taken as the volume form ${1\o 3!} J\w J\w J\,$.  This differs by a minus sign with the convention used in \GidKP.}  Equation \heqb\ characterizes the required type of
the compactification metric.  It specifies that there must exist on any
$N=1$ background geometry a universal closed $(2,2)$-form,
$e^{-2\phi}(J\w J)$.  In particular, by making the conformal
rescaling, $J=e^\p \tJ$,  we obtain 
\eqn\heqbal{\rd(\tJ\w\tJ) =0~, }
which is the defining condition for a balanced metric \Michel.  Thus,
we arrive at the statement that a supersymmetric heterotic flux
compactification requires a conformally balanced metric.   

An interesting and obvious subset of the balanced metrics are those
that satisfy additionally the K\"ahler condition, i.e. $\rd \tJ=0$.
For backgrounds with conformally K\"ahler metrics, the H-flux has the
property from \Hconda\ and \Hnp\ that 
\eqn\confK{H=i(\bpa-\pa)(e^\phi \tJ) = i(\bpa-\pa)\phi\w (e^\phi \tJ)
= H^{NP}~.} 
which implies that the primitive part of the H-flux, $H^P$, is zero.
Indeed, for backgrounds with $H^P=0$, the compactification metric is
conformally K\"ahler.  With also the presence of the holomorphic
3-form and SU(3) holonomy, the underlying manifold is actually a
Calabi-Yau manifold though the Ricci-flat metric is not the physically
relevant metric here.   

In summary, we see that the $N=1$ background manifold is required to
be hermitian with zero first Chern class.  The metric types can be
characterized by the H-flux as follows.  
\eqn\table{\matrix{H= & H^{NP}+H^P\quad\quad\quad &{\rm conformally\ balanced\
metric}\cr H =& H^{NP}\qquad\qquad\qquad &{\rm conformally\
Kahler\ metric}\quad\! \cr H= & 0 \qquad\qquad\qquad\quad~ &{\rm
Calabi-Yau\ metric}\qquad\quad~\, } }

\subsec{Deformations of heterotic solutions}

For a heterotic flux solution, it is phenomenologically important to
know the moduli space of the solution with $N=1$ supersymmetry.  In
terms of the four-dimensional low energy effective theory, each
modulus corresponds to a massless scalar field.  Given a solution, the
moduli space can be found by deforming the solution such that it
continues to satisfy the $N=1$ supersymmetry conditions discussed above.  

There is at least one modulus in all heterotic supergravity
compactifications at the classical level.  This is the
dilaton modulus \Strom\ arising from shifting the dilaton by a
constant, $\p\to \p +c $, and keeping all other fields fixed.  Indeed,
the supersymmetry conditions and also the equations of motion are
invariant under a constant dilaton shift. 

Besides the dilaton modulus, a generic compactification will have other
moduli from varying the other fields in the theory.  We will write down
the conditions for the existence of moduli associated with the
metric. To begin, recall that the metric satisfy the hermitian condition
\eqn\hermcond{g_{mn}=J_m{}^rJ_n{}^sg_{rs}~.}
In varying the metric, the hermitian condition can require a
corresponding variation in the complex structure.  However, the
complex structure can not be deformed arbitrarily \GSW.  Starting from the
standard complex structure, $J_a{}^b=i\d_a{}^b$ and
$J_\ba{}^\bb=-i\d_\ba{}^\bb$, consider the deformation 
\eqn\compdef{J_m{}^n\to {J'}_m{}^n = J_m{}^n + \t_m{}^n~.}
The condition ${J'}_m{}^r{J'}_r{}^n=-\d_m{}^n$ implies that the
only non-zero components are $\t_\ba{}^b$ and their complex conjugates
$\t_a{}^\bb$.   Furthermore, imposing the vanishing of the Nijenhuis
tensor leads to the requirement that $\t_\ba{}^b$ are elements of
$H^1(T)$, the first
Dolbeault cohomology of the internal manifold $K$ with values in the
holomorphic tangent bundle $T$.  With the $SU(3)$ holonomy, we can use
the holomorphic three-from, $w_{abc}$, to construct the associated
$(2,1)$-forms, $\chi_{\ba b c}=\t_\ba{}^a w_{abc}\,$, which are $\bpa$ closed
and elements of $H^{(2,1)}$.  Therefore, the dimension of the complex
structure deformation on any complex manifold with SU(3) holonomy is
given by the Hodge number $h^{(2,1)}$. 

Now varying the hermitian condition \hermcond\ to linear order, we find that 
variations of the complex structure, $\d J_m{}^n = \t_m{}^n$ with
$\t_\ba{}^b\in H^1(T)\,$, and the metric
variations of pure components are linked as follows
\eqn\metvarc{\d g_{ab}= {i\o 2}\, (\t_a{}^\bc g_{b \bc} + \t_b{}^\bc
g_{a \bc})~,}
and similarly for the complex conjugates $\d g_{\ba\bb}$.\foot{Since
all fields are real, we will not explicitly write out components of
fields which are just complex conjugates of other components.}  
We can therefore divide metric variations into two types.  The pure
components, $\d g_{ab}$ and $\d g_{\ba\bb}\,$, are associated with the
deformations of the complex structure, while the mixed components, $\d
g_{a\bb}\,$, are independent of the complex structure.  The mixed
components however are associated
with the variation of the hermitian form.  With $J_{mn}=J_m{}^k
g_{kn}\,$, the variation of the hermitian form is given by
\eqn\hermva{\eqalign{\d J_{ab} &= \t_a{}^\bc g_{b \bc} + i\,\d g_{ab}= {1\o 2}\, (\t_a{}^\bc g_{b \bc} - \t_b{}^\bc
g_{a \bc})~, \cr \d J_{a \bb}&= i\, \d g_{a\bb}~.}}

We now perform a linearized variation on the supersymmetry
constraint equations.  A variation in the metric will generically
require a variation in both the H-flux and the dilaton.  This can be seen
in the variation of the two constraint equations, \Hcondaa\ and
\Hcondb, which we re-write here in variational form
\eqn\Hcondav{\d H_{mnp}= 3\,\d({J_{[m}}^r\nabla_{|r|}J_{np]})~,}
\eqn\dilatonv{(\rd\, \d\p)_m = {1\o 4}\, \d(J_m{}^n H_{nrs} J^{rs})~.}
Recalling that $H=i(\bpa-\pa)J\,$, we see that the right hand side of
both equations above depend only on the metric and its variation.  Therefore,
\Hcondav\ and \dilatonv\ are the determining equations for the
variations of the H-flux and the dilaton, respectively.  Conversely,
any condition on the variations of the H-flux or the dilaton will
reflect back as a constraint for the metric variation.  From
the form of equation \dilatonv, a constraint for metric variations is
immediately apparent.  We see that a solution for $\d\p$ exists if and
only if the one-form
$A_m=\d ({J_m}^n\, H_{nrs}\, J^{rs})$ is exact.  (If we treat $A_m$ as
a U(1) gauge field, then we have the pure gauge condition,
$A=\rd\Lambda$.)  The trivial integrability condition $\rd^2 \d\p=0$ gives
a non-trivial metric variational constraint
\eqn\integcond{(\rd A)_{mn}= 2\,\pa_{[m}\, \d (J_{n]}{}^p\, H_{prs}\,
J^{rs})=0~.}

Now more explicitly, \Hcondav\ gives the H-flux variation
\eqn\Hthreev{\d H_{abc} = {i\o 2} (\t_a{}^\bd H_{\bd bc} + \t_b{}^\bd
H_{a \bd c} + \t_c{}^\bd  H_{a b \bd } )~.}
\eqn\Htov{\d H_{ab\bc} = - 3i\, \pa_{[a\,} \d J_{b\bc\,]}  + i (
\t_a{}^\bd H_{\bd b \bc} + \t_b{}^\bd H_{a\bd \bc})~.}
The equation for $\d H_{abc}$ can also be obtained by requiring that
the $(3,0)$ component of $H$ with respect to the deformed complex
structure remains zero.  And also, we obtain from \dilatonv\ 
\eqn\dilatv{\pa_a \d \phi = {1\o 2}\,\t_a{}^\bd H_{\bd b \bc}\,
J^{b\bc} + {i\o 4}(2\,
H_{ab\bc}\,\d J^{b\bc} + 2\, \d H_{ab\bc}\, J^{b\bc} + H_{a\bb\bc}\,\d
J^{\bb\bc}) } 
with $\d J^{\bb\bc}={1\o 2} \,(g^{\bb d}\t_d{}^\bc- g^{\bc d}
\t_d{}^\bb)$.  The above three equations, \Hthreev, \Htov, and
\dilatv, represent necessary variational conditions relating the
metric, H-flux, and the dilaton.

The above variational conditions can easily reproduce the
metric moduli space for the zero-flux Calabi-Yau solution.   With $H=0$, the
variational equations for $\d H$ and $\d\p$ above trivialize and
become independent of
$\t_m{}^n$.  Therefore, each of the $h^{(2,1)}$ complex structure
deformations with $\t_\ba{}^b\in H^1(T)$ represents a modulus.
Furthermore, requiring $\d H=0$ so that the solution remains
Calabi-Yau, \Htov\ implies that $\rd( \d J) =0$, which is a
variation on the K\"ahler condition,  $\rd J=0\,$, and has $h^{(1,1)}$
independent solutions.  (Those $\d J\,$ which are exact correspond to
diffeomorphism and are thus modded out.)  Altogether for the zero-flux
Calabi-Yau, we have $h^{(2,1)} + h^{(1,1)}$ independent metric
variations that do not involve either the H-flux or the dilaton field.

When the H-flux is non-zero, characterizing all metric variations that
satisfy even just the metric variational constraint \integcond\
seems difficult.  To proceed further,
we will make a simplifying restriction by taking the H-flux to be
fixed, or equivalently, imposing $\d H =0$.  Fixing the H-flux will typically
also fix the complex structure.  From \Hthreev, we have the constraint
on complex structure that
\eqn\Hthreev{\d H_{abc} = {i\o 2} (\t_a{}^\bd H_{\bd bc} + \t_b{}^\bd
H_{a \bd c} + \t_c{}^\bd  H_{a b \bd } )=0~.}
This can be a strong constraint since both $\t_m{}^n$ and $H$ are
generically functions and are directly dependent on the geometry.
Moreover, if there are regions of the manifold where $\rd H=0$ (as is
the case for the orbifold background we discuss in the next section), then 
the integral of $H$ over three-cycles in these regions must be quantized
\dirac.   Though the integral of $H$ is independent of the
hermitian metric, it can depend on the complex structure via the
limits of integration for the coordinates.  Specifically, the
boundary conditions for the coordinates on a compact manifold can be
modified by a deformation of the complex structure.\foot{Consider the
simple example  of a torus.  The complex structure is parametrized by
the modular parameter $\t=\t_1+i\t_2$.  The metric can be written as
$ds^2= A\, dz d{\bar z}$ where $A$ is the area of the torus and
$z={1\o \sqrt{\t_2}} (x+\t y)$ with $x\sim x+1$ and $y\sim y+1$.
Clearly, the boundary conditions (or the periodicities) of $z$ depend
on $\t$.}  Hence, imposing the Dirac quantization in regions where $\rd
H=0$ can also fix the complex structure.  Below, we shall analyze
variations in a fixed H-flux background and assume that the complex
structure has been fixed.

With $\d H=\t_m{}^n=0$, the two conditions, \Htov\ and
\dilatv, and their complex conjugates can be expressed as 
\eqn\vheta{\d H =i(\bpa-\pa)\,\d J =0~,}
\eqn\vhetb{(\rd \d\p)_m = {1\o 4} {J_m}^n\, H_{nrs}\,\d J^{rs}~.}
where $\d J$ denotes the variation of the mixed components of the
hermitian form, $\d J_{a\bb}\,$.  The two equations can also be
obtained directly by
varying equations \Hconda\ and \Hcondb.  Interestingly, in a fixed
H-flux background, though $\rd
J\neq0$, \vheta\ implies that $\rd (\d J)=0$ or that the variation of
the hermitian form is required to be a closed
$(1,1)$-form.  Equation \vhetb\ is the determining equation for the
variation of the dilaton.  The
associated integrability condition gives the second constraint for $\d J$ 
\eqn\vhetc{(\rd A)_{mn} = 2 \,\pa_{[m}\left({J_{n]}}^p\, H_{prs}\,\d
J^{rs}\right)=0~.}
This second constraint is non-trivial and involves the fixed H-flux.
A valid deformation of $\d J$ must satisfy both of the above conditions,
\vheta\ and \vhetc.

Together, the above two conditions for $\d J$ are
not easily satisfied.  Consider first an exact variation, $\d J= i
\pa\bpa\, \Phi$.  Being exact, it is trivially a closed (1,1)-form.
However, an exact variation of $J$ typically does not satisfy \vhetc.
(We will give an explicit example of this in the next section.)
Therefore, although $\d J$ is closed, the moduli are not related to
the standard de Rham cohomology.  As a second example, consider the
radial modulus $\d J_{a\bb} =
aJ_{a\bb}$ which is present in Calabi-Yau compactifications and
represents an
overall rescaling of the internal metric.  A variation of $\d\p= \p/a$
will satisfy \vhetb\ (and hence also \vhetc) but now \vheta\ is not
satisfied.  A radial modulus would  give $\d H= i(\bpa-\pa)\d J = a
H\neq 0 $ which is not allowed.  The essential point here is that $J$
is not a closed (1,1)-form for non-K\"ahler flux backgrounds.  We thus
see that fixing the H-flux also fixes the overall scale of the metric.

\newsec{Moduli of the heterotic solution dual to type IIB orientifold}

The $N=1$ heterotic solution with H-flux discussed in
\refs{\DasRS,\BD,\BBDG,\BBDGS} is dual to type IIB $N=1$ flux
compactification on an orientifold.  Without taking account of
non-perturbative effects, the IIB orientifold models with fluxes
lift all complex structure moduli but leave unfixed certain K\"ahler
moduli \refs{\KacST,\TriT}.  This suggests at the supergravity level,
the heterotic dual also contain moduli.  In this section, we find the
heterotic moduli via duality and show that they satisfy the conditions
discussed in the previous section.

\subsec{IIB orientifold model and its moduli}

We first review the type IIB flux model on a toroidal orientifold
described in \refs{\DasRS,\BD, \BBDG}.  The background geometry is
$K3 \times T^2/\IZ_2$ where $\IZ_2=\O(-1)^{F_L}\CI_{2}$ and $\CI_{2}$ is the reflection operator reversing the two directions of $T^2$.\foot{The $\O$
operator here refers to the worldsheet parity operator and should not be
confused with the holomorphic three-form in section 2.}  The solution
originates from an M-theory (or dual F-theory) flux compactification
on $K3\times K3$.  Localized at each of the four fixed points of
$T^2/\IZ_2$ are 
four D7-branes and an O7-plane.  For concreteness, the
orbifold limit of $K3$ is taken, replacing $K3$ with $T^4/\CI_4$.  This
allows us to write down a warped metric of the form
\eqn\mwarped{ds^2 = \D^{-1}\, \eta_{\m\n}\,dx^\m dx^\n + \D\, g_{mn}\, dy^m
dy^n}
where $g_{mn}$ is the metric on $T^4/\CI_4\times T^2/\IZ_2$ and the
warped factor $\D$ is related to the five-form field strength
\eqn\fff{\eqalign{F_5 &= d D_4 - {1\o 2} C_2 \w H_3  + {1\o 2} B_2 \w
F_3 \cr &= (1+ *_{10})\, (d\D^2 \w dx^0 \w dx^1 \w dx^2 \w dx^3)~.}}
Here, the Hodge star, $*_{10}$, is defined with respect to the ten-dimensional
metric.  The self-duality of $F_5$ further implies
\eqn\wwflux{*_{6}\, d\D^2 =  - {1\o 2}  C_2 \w H_3  + {1\o 2} B_2 \w F_3 }
where $*_{6}$ is defined with respect to the unwarped six-dimensional
metric, $g_{mn}$ in \mwarped.  This condition relates the warp
factor to the internal metric and the fluxes.

In order to preserve $N=1$ supersymmetry, the three-form flux, $G_3$, is
required to be a primitive (2,1)-form \GraPol.  For a diagonal metric with $J = i g_{1\aa}\,\dza\w\dbza + i g_{2\ab}\,\dzb\w\dbzb + i g_{3\ac}\,\dzc\w\dbzc\,$, the constant
non-localized part takes the form\foot{Additionally, $G_3$ can have a metric dependent term proportional to $(g_{1\aa}\, \dza\w\dbza - g_{2\ab}\, \dzb\w\dbzb)\w \dzc\, $ which we do not consider here for simplicity.}
\eqn\Gflux{\eqalign{G_3 &= F_3 - \t H_3 \cr &= -2i~
{\rm Im}[\t] (\bA\dza\w\dbzb\w\dzc + \bB\dbza\w\dzb\w\dzc + \bC
\dza\w\dzb\w\dbzc )}} 
with $\t=C_0 + i\,e^{-\p_B}$ and
\eqn\Hval{\eqalign{H_3&= A\dbza\w\dzb\w\dbzc + B \dza\w\dbzb\w\dbzc +
C\dbza\w\dbzb\w\dzc +  ~{\rm c.\ c.} \cr F_3 &= \t ( A\dbza\w\dzb\w\dbzc + B
\dza\w\dbzb\w\dbzc + C\dbza\w\dbzb\w\dzc) + ~{\rm c.\ c.}~~. }}
where $A,B$ and $C$ are complex constants and ``c.~c.'' denotes
the complex conjugates.  Above, the $z_i$'s are the holomorphic coordinates
of the internal six-manifold.  We will choose the associated two-form
potentials to be 
\eqn\Bval{\eqalign{B_2&=\left[(A\bza + {\bC\o 2}\za)\dzb - (B\bzb +
{\bC\o 2}\zb)\dza\right] \w \dbzc + ~{\rm c.\ c.} \cr C_2&=
\left[(\t A\bza + \bt{\bC\o 2}  \za ) \dzb
- (\t B\bzb + \bt{\bC\o 2}\zb) \dza\right] \w\dbzc +  ~{\rm c.\ c.}~~. }}
Additionally, the three-form fluxes are constrained by the
D3-brane tadpole cancellation condition
\eqn\tadp{{\chi(K3\times K3)\o 24} = -\int_K\, H_3\w F_3 + N_{D3}~,} 
where $N_{D3}$ is the number of D3-branes present and we have set
$(2\pi)^2\a'=1$ in line with \refs{\DasRS, \BD}.  The three-form
fluxes are also quantized,
\eqn\hfquan{\int_\G\, H_3 \in \IZ~, \qquad\qquad \int_\G\, F_3 \in 
\IZ~,}
where the integration is over any three-cycles.

We now consider the moduli of this type IIB orientifold model.  We
start with the complex structure moduli.  The
condition that $G_3$ is a (2,1)-form can either fix the complex
structure moduli and $\t$ when a set flux is turned on, or determine the
allowed flux for a given complex structure and $\t$.  For ease of
performing duality, it is more useful to set the complex structure
by defining the holomorphic variables $z_1 = y_4 +i y_5\,,~
z_2=y_6 + i y_7\,,~z_3 =y_8 +iy_9\,$ (with the real coordinates having
unit periodicity on the covering space) and also set
\eqn\tauf{\t=C_0+ i\,e^{-\p_B} =\, i~.}
The additional constraints on the fluxes, \tadp\ and \hfquan, then become
\eqn\tadpr{24= 4\left(|A|^2+|C|^2+|B|^2\right)+N_{D3}}
\eqn\hfquanr{{\rm Re\,}A\pm{\rm Re\,}B\pm{\rm Re\,}C\in 2\IZ~,\qquad {\rm
Im\,}A\pm{\rm Im\,}B\pm{\rm Im\,}C\in 2\IZ~.}
These fix the fluxes to discrete values of $(A,B,C)$.  Some examples
with $N_{D3}=0$ are for instance 
$(A,B,C)=(2+i,i,0)$~\refs{\BD,\BBDG}, $(A,B,C)=(2,1,1)$ \DasRS\ and
also $(A,B,C)=(1,1,2)$ where $A=B$.  

With fixed complex structure and axion/dilaton, we turn
now to the K\"ahler moduli.  Recall that the K\"ahler moduli arise
from varying the K\"ahler form $J\to J + \d J$
which changes the metric via the relation $J_{a\bb} = i g_{a\bb}$.
When $G_3=0$, the number of K\"ahler moduli is given by $h^{1,1}$.
For a non-zero fixed  $G_3$, the supersymmetry condition that $G_3$ is
primitive will lift some of the K\"ahler moduli \KacST.  For a
middle-dimensional form such as $G_3$ (i.e. a 3-form in six dimensions), the
primitivity condition is equivalent to the constraint $G_3\w J =0$.  
The unlifted K\"ahler moduli are those that satisfy
\eqn\Kcond{G_3 \w \d J =0~.}
This ensures that $G_3$ remains primitive and its self-duality
property $*G_3=-iG_3$ unchanged with respect to the deformed
metric.\foot{In our convention for the Hodge star, the equation of motion 
requires $G_3$ to be imaginary anti-self-dual.  Note that all deformations
satisfying \Kcond\ will preserve the imaginary anti-self-duality
condition of $G_3$ since any primitive $(p,q)$-form for $p+q=3$ satisfies
$*A^{(p,q)}=i(-1)^q A^{(p,q)}$.  Also, as noted in
\GidM, those K\"ahler moduli which do not satisfy \Kcond\ may still
correspond to flat directions if we allow $G_3$ to vary.  However,
here $G_3$ is kept
fixed so those moduli not satisfying \Kcond\ are lifted.}  For generic
values of $(A,B,C)$, the K\"ahler deformations that satisfy \Kcond\ are 
\eqn\Ja{\d J=i\, \d k_1\, \dza\w\dbza + i\, \d k_2\, \dzb\w\dbzb + i\, \d k_3\, \dzc\w\dbzc~,}
For special values of $(A,B,C)$ such that $A=\pm B$, an additional
deformation is present.  Consider for example, $(A,B,C)=(1,1,2)$, the
deformation of 
\eqn\Jb{\d J=i\, \d k_4 (\dza\w\dbzb + \dzb\w\dbza)}
is also allowed.  Note that as required, these K\"ahler deformations
are real and invariant under the orientifold action.  

We will incorporate the K\"ahler deformations directly into the
internal six-dimensional metric.  We promote the deformation
variables $\d k_i$ into parameters of the metric and write the metric as follows
\eqn\metric{ds^2 = \D\,g_{mn}\,dy^m dy^n = 2 \,\D\, g_{a\bb}\, dz^a dz^\bb  }
with
\eqn\mvar{g_{a\bb}=\left(\matrix{k_1 & k_4 & 0\cr k_4 & k_2 & 0 \cr 0
& 0 & k_3 } \right)~. }
The $k_i$'s, for $i=1,2,3\,$, parametrizes the volume of each
torus.  And $k_4$ which is present only if $A=B$, mixes the two tori
of $T^4/\CI_4$.\foot{We have included the $k_4$ parameter to illustrate that any deformation that arises for special values of $(A,B,C)$ can be incorporated into the duality mapping performed below.}  With explicit expressions for both the metric and
3-form fluxes, we can now determine the warp factor using the
condition for the self-duality of $F_5$ in \wwflux.  The warp factor
satisfies two independent differential equations 
\eqn\fluxea{k_3 \left(k_2 \pa_\bza - k_4 \pa_\bzb\right) \D^2  = -
\Big(|A|^2+{1\o 2}|C|^2\Big)\, \za~,}
\eqn\fluxeb{k_3 \left(k_1 \pa_\bzb - k_4 \pa_\bza\right) \D^2  = -
\Big(|B|^2+{1\o 2}|C|^2\Big)\, \zb~,}
which has the solution  
\eqn\warpf{\eqalign{k_3\,\D^2 = c_0 -{1\o k_1k_2-k_4^2}\,\bigg\{
k_1\Big(|A|^2&+{1\o 2}|C|^2|\Big)|\za|^2 + 
k_2 \Big(|B|^2+{1\o 2}|C|^2\Big) |\zb|^2 +\cr & +{1 \o 2} k_4
\Big(|A|^2 + |B|^2 + 
|C|^2\Big)(\za\bzb + \zb\bza)~\bigg\}~,}}
where $c_0$ is an integration constant which ensures that
$\D^2$ is positive definite.  It is worthwhile to point out that the warp
factor has dependence on the K\"ahler moduli.  This has also been noted in
\refs{\GidM}.  

We have obtained the unlifted K\"ahler moduli and will proceed
to dualize the moduli in the next subsection.  But before doing
so, it is important to point out that we have thus far ignored the
contributions from the localized fixed points.  The 
localized fixed points generate singularities for the warp factor.
They also allow for localized $G_3$ fluxes corresponding to gauge
fluxes on the D7-branes.  These two issues do play an important role
in the consistency of the theory and have been worked out in detailed in
\refs{\BD,\BBDG}.  Moreover, if we treat $T^4/{\cal I}_4$ as
a true $K3$ surface, then we should also consider the metric
deformations at the orbifold fixed points.  To see that they are
present, recall first that $K3$ has $58$ metric deformations \Wal.  Of
these, only $10$ are due to the deformation of the $T^4$ metric.  The
rest, $48=3\times 16\,$, comes from the $16$ fixed points which must
be blown up and each replaced with an Eguchi-Hanson space.  Of the
three metric deformations of the Eguchi-Hanson space, one is an
overall scaling and two are rotations \Page.  Some of the moduli at the
fixed points can be fixed by the localized $G_3$ flux but not by the
non-localized $G_3$ flux in \Gflux.  However, since the duality
mapping that we will perform in the next subsection is only at
the level of supergravity, we will focus only on the non-localized
moduli and not consider further the localized moduli at the singularities.

\subsec{Dual heterotic model and its moduli}

Without fluxes, the type IIB orientifold model on $K3 \times
T^2/\IZ_2$ becomes the type IIB theory on $K3 \times T^2/\O$ after applying
T-duality in both directions of $T^2$.  This is equivalent to the type
I theory on $K3\times T^2$, which in turn is S-dual to the SO(32) heterotic
theory on $K3\times T^2$.  The same duality chasing can be worked out
with the presence of fluxes for the IIB orientifold model on
$T^4/\CI_4 \times T^2/\IZ_2$.  Note that all type IIB fields in
the previous subsection do not have any dependence on $\zc, \bzc$,
which are the directions which we T-dualize.  The transformation rules
of supergravity fields under T- and S-dualities are derived for example in
\refs{\BergHO,\MesO,\Hassan}.  Applying the dualities, the heterotic dual
has a $M^{3,1}\times K$ background geometry, and the dilaton, internal
metric, and B-field are given as follows\foot{Except for the type IIB
potentials, $B_2$ and $C_2$,
all other background fields in this subsection are heterotic fields.}
\eqn\gH{e^{\p}= 2\, \D\; k_3~,}
\eqn\metH{ds^2 ={ e^{2\p}\o k_3}\big( k_1\, \dza\dbza +  k_4\,
(\dza\dbzb + \dzb\dbza)+ k_2 \,\dzb\dbzb \big) +|\dzc + 2\,
(B_2)_{a\bzc} dz^a|^2~,}
\eqn\BHB{B_{mn} = i(B_2\w C_2)_{mn \zc\bzc}~,\quad
 B_{m\zc}=-i(C_2)_{m\zc}~,\quad B_{m\bzc}=i(C_2)_{m\bzc}~.}
Substituting  the expressions for $B_2$ and $C_2$ in \Bval, we have
more explicitly,
\eqn\metm{\eqalign{&g_{a\bb} =\cr
&{1\o 2}\left(\matrix{e^{2\phi}{k_1\o k_3}+|2B\bzb+\bC\zb|^2  &
e^{2\phi}{k_4\o k_3}\!-\!(2B\bzb\!\!+\!\bC\zb\!)(2\bA\za\!\!+\!C\bza\!) &
-2B\bzb \!-\! \bC\zb\cr e^{2\phi}{k_4\o
k_3}\!-\!(2\bB\zb\!\!+\!C\bzb\!)(2A\bza\!\!+\!\bC\za\!)&
e^{2\phi}{k_2\o k_3}+|2A\bza+\bC\za|^2 & 2A\bza + 
\bC\za\cr -2\bB\zb\!-\!C\bzb & 2\bA\za+C\bza & 1}\right)}}
\eqn\BI{\eqalign{B&=(\bB\bC\zb^2-BC\bzb^2)\,\dza\w \dbza +
(\bA\bC\,\za^2 - AC \,\bza^2)\,\dzb\w\dbzb~ + \cr &+
 \bigg[( BC \,\bza\bzb-\bA\bC\za\zb)\, \dza\w\dbzb +  
\big((\bB\zb\!-\!{C\o 2}\bzb)\, \dbza\!-\!(\bA\za\!-\! {C\o 2} \bza )\, \dbzb
 \big) \w \dzc+~ {\rm c.\ c.}\bigg]}} 

We see from \metH\ that the internal space consists of a warped
$K3$ (or more precisely $T^4/\CI_4$) with a non-trivial $T^2$ fiber. 
The twisting of the $T^2$ fiber arises from the two T-duality
transformations and is due solely to the presence of the B-field in
the IIB orientifold model \KacSTT. 
Further insight can be obtained by decomposing this six-dimensional
compactification as a four-dimensional compactification plus a 
two-dimensional one.  First, in compactifying down to six dimensions, 6d
$N=1$ supersymmetry requires that the compact four-manifold be a
conformal Calabi-Yau with the warp factor $e^{2\phi}$ \refs{\GauMW,
\BT}.  Here, we have a conformal $K3$ with exactly the warping mandated
by supersymmetry.  Now if we proceed further to compactify on a $T^2$,
we will obtain a background with 4d $N=2$ supersymmetry.  Thus, the
presence of the non-trivial $T^2$ fibering here is required to break the
additional supersymmetries so that we end up with the desired $N=1$
supersymmetry in four dimensions.

The dual heterotic background can be checked to satisfy the
heterotic supersymmetry constraints \Hconda\ and \Hcondb.  The metric
\metm\ is in fact conformally balanced.  Using $H=
i(\bpa-\pa)J$, the non-zero H-field components (modulo complex
conjugation) are
\eqn\Hval{H_{\aa\ab2}= 3\,AC\bza~,\quad H_{\aa1\ab}=3\,BC\bzb~,\quad
H_{13\ab}=\bA~,\quad H_{32\aa}=\bB~,\quad H_{12\ac}=\bC~.}
We note that the contributions to the H-field come from both the warp
factor and the $T^2$ fiber part of the metric.  We have however used
the warp factor relations, \fluxea\ and \fluxeb, to cancel off
additional terms in the first two components above.  The
$H$ values agree with those obtained from $H=\rd B$ using \BI.  And
for $C\neq 0$, $H_{\aa\ab2}$ and $H_{\aa1\ab}$, like the B-field, are only
locally defined on patches.  Moreover, it is important to point out
that we have $\rd H=0$ here only because the contributions localized
at the fixed points of the background geometry have not been taken
into account.  The curvature singularities at the fixed points of
$T^4/\CI_4$ lead to additional contributions for the H-field through
the warp factor \BBDG.   Indeed, from a no-go theorem, $\rd H$ can not
vanish in a non-zero flux background \refs{\deWSH,\IvaP}.

The moduli of the heterotic model can be simply obtained by varying the
$k_i$'s in the expressions for the background fields in \gH, \metH,
and \Hval.  We find that the variations only
deform the dilaton field and the conformal $T^4/\CI_4$ part of the
metric.  The H-flux given in \Hval\ has no dependence on the $k_i$'s
and is thus fixed.  The complex structure of the heterotic model
is also fixed in the fixed H-flux background.

Let us now consider the variations in more detail.  Take first the $k_3$
variation.  In the IIB orientifold model, $k_3$ varies the volume of
$T^2/\IZ_2$.  In contrast, the heterotic dual metric has no dependence 
on $k_3$ at all.  The factor $e^{2\p}/k_3=4k_3\D^2$ in the metric
\metH, as given explicitly in \warpf, is $k_3$ invariant.  $k_3$ only
shifts the dilaton and is thus the heterotic dilaton modulus.  As for
$k_1, k_2, k_4$, the remaining
non-localized K\"ahler moduli of $T^4/\CI_4$ in the IIB orientifold model,
we find that their variations in the heterotic model deform both
$\phi$ and $g_{a\bb}$.  Take for example $k_1$.  Its variation results
in the following field deformations
\eqn\kmodud{\d_{k_1} \p = {1\o 2 k_1^2} {(|B|^2+{1\o 2}|C|^2)|z_2|^2\o
k_3\D^2}\,\d k_1~,}
\eqn\kmodug{\d_{k_1} g_{a\bb} = \left(\matrix{2 ( c_0-{1\o
k_2}(|A|^2+{1\o 2}|C|^2)|z_1|^2) & 0 & 0\cr 0& 2 {k_2 \o
k^2_1}(|B|^2+{1\o 2}|C|^2)|z_2|^2&0\cr 0& 0& 0}\right) \d k_1~,}
where we have set $k_4=0$ above to simplify the
expressions.\foot{Recall that the $k_4$ deformation is present only if
$A=B$.  A non-zero $k_4$ will add more terms to the expressions.}  The
non-zero variation for $g_{2\ab}$ comes from the $k_1$
dependence of the warp factor.  ($\d g_{1\ab}$ and $\d g_{2\aa}$
are also non-zero if $k_4\neq 0$.)  As is evident, the simple  K\"ahler
moduli in the IIB orientifold model have been mapped to non-trivial
deformations in the heterotic dual.  Nevertheless, the deformations
can be straightforwardly checked to satisfy both $\rd (\d
J)=0$ \vheta\  and the determining equation for $\d\p$ in \vhetb.
Thus we have via duality provided explicit variations that satisfy the
two moduli conditions given in section 2.2 for a fixed H-flux background.

The explicit metric moduli that we have found are in fact very special.
Indeed, it is not easy to write down metric variations that satisfy
both variational constraints.  Even for variations $\d J$ that are
exact and hence trivially  closed, the second condition, requiring the
1-form $A_m= {J_m}^n\,
H_{nrs}\,\d J^{rs}$ also be exact, is not automatically satisfied.  We
demonstrate this 
with an explicit example.  For simplicity, we take $k_1=k_2=k_3=1$ and
$k_4=C=0$.  To construct the exact 2-form, we 
utilize the natural 1-form on the $T^2$ fiber, $\a=  \dzc +
2(B_2)_{a\bzc} dz^a= \dzc + 2A \bza \dzb - 2 B\bzb \dza$.  We take 
\eqn\djex{\d J = \e\, \rd( \a +  {\bar \a})=2\e (\bA+B) \dza \w \dbzb
- 2\e (A + \bB) \dzb\w\dbza~.  } 
where $\e$ is the infinitesimal parameter.  The values of $A_m$ can
now be straightforwardly worked out with the H-flux given in \Hval.
We find 
\eqn\Aform{\eqalign{ A_{\za} & =-8i\e\,e^{-4\p}\bA B(A+\bB)\bzb \cr
A_{\zb} &= 8i\e\, e^{-4\p} A\bB(\bA+B)\bza \cr A_{\zc} & =4i\e\,
e^{-4\p}(|A|^2+|B|^2+2\bA\bB)}} 
Clearly, $A_m$ has no dependence on $\zc$ and $\bzc$ since no fields
in the heterotic dual model has dependence in the coordinates which
were T-dualized.  But as can be easily checked, $\pa_{\za} A_{\zc}$
and $\pa_{\zb} A_{\zc}$ are non-zero, and therefore,
$(dA)_{\za\zc}$ and $(dA)_{\zb\zc}$ do not vanish.  We thus conclude
that $A_m$ is not exact and $\d J$ as given in \djex\ is not a valid
deformation.  Hence, we have shown that a variation of $\d J$ by an
exact form does not generically give a new solution. 

Finally, it is significant that the $T^2$ fiber is not affected by any
of the $k_i$ variations.  With $2\pi\sqrt{\a'}=1$, the area of the
$T^2$ fiber is fixed and is of order $\a'$ \refs{\BecBDP,\BBDGS}. 
Therefore, there is no overall radial modulus in the heterotic model,
agreeing with our analysis in section 2.2.  This is as expected since the
heterotic dual backgrounds can not have a continuous deformation to a large
radius Calabi-Yau geometry \BD.   However, the small size of the $T^2$
fiber may point to the need for incorporating $\a'$ corrections into
the analysis.  But with regard to the metric moduli that we 
have found via duality, it is useful to emphasize that these moduli
only involve the conformal $T^4$ base and does not affect the small
directions of the geometry.  The $K3$ base can be taken to be
sufficiently large such that the supergravity analysis on the
conformal $T^4/\CI_4$ remain valid.  A large $T^4/\CI_4$ base is in
fact also required in order to neglect the contributions from the
fixed-point singularities.  Heuristically, we can think of the fixed
points as located at the ``boundaries'' of $T^4/\CI_4$ and we have
studied the moduli in the ``bulk,'' far away from the fixed points.

\newsec{Discussion}

We have analyzed the linearized variational equations of the
supersymmetry constraint equations to study moduli associated with the
metrics.  In a fixed H-flux background and with the complex
structure fixed, we wrote down two conditions
\vheta\ and \vhetc, which metric variations must satisfy.  Our
analysis in the heterotic theory is also directly applicable to the
type I theory and to the subsector of type II theories where only the
H-flux is turned on. 

Our focus on the metric moduli in a fixed H-flux background is
primarily motivated by a desire to understand the heterotic dual of the
unlifted K\"ahler moduli in certain type IIB orientifold models.
However, in order to understand the full heterotic moduli space, it is
necessary as the next step to consider more general deformations.
Generically, we expect there can be moduli associated with the variation of
the complex structure if the H-flux is allowed to vary.  Another
source of moduli can come from the gauge fields.  From the gaugino 
supersymmetry transformation, the gauge field strengths $F_{mn}$ are
required to satisfy the Hermitian Yang-Mills equations,
i.e. $F_{ab}=F_{\ba\bb}=0$ and $g^{a\bb}F_{a\bb}=0$.  To our
knowledge, not much is known about the space of gauge field solutions in
conformally balanced geometries.  (Recently, progress in
constructing non-Abelian gauge field configurations in non-K\"ahler
geometries has appeared in \Yau.)  Note that the gauge field strengths, the
metric, and the H-flux are all connected via the modified Bianchi
identity \hetd.  Thus, it would be interesting to understand if a
variation in the gauge field can be independent or would also require
a variation in the metric and the H-flux.  And conversely, whether a
metric variation that lead to a variation of ${{\rm tr}\,( R^{(-)} \w
R^{(-)})}$ can always be accompanied by a compensating variation in
${{\rm tr}\,( F \w F)}$ in  backgrounds with $dH\neq 0$.  These issues are
important and worth pursuing.

Nevertheless, knowing certain required conditions for the metric
moduli in a fixed flux background, it is interesting to ask whether
there is a simple method to count the number of moduli satisfying these
conditions.  For instance, on a Calabi-Yau, the equation $\rd (\d J)=0$
is the condition for the K\"ahler moduli and the number of moduli is given
simply by the Hodge number $h^{(1,1)}$.  So perhaps the number of metric
moduli can also be simply expressed in terms of the dimension of a generalized
cohomology.  We briefly explore this issue below \work. 

Consider first the IIB case.  The unlifted K\"ahler moduli must
satisfy the two constraints
\eqn\IIBm{ \rd\, \d J = 0~~,\qquad\qquad G_3 \w \d J =0~.} 
Notice that the two constraints each separately defines a cohomology,
i.e. $d^2=0$ and also $G\w G\w =0$.  But since we are interested in
constructing one 
cohomology, consider combining the two together as
\eqn\IIBma{\CD\,\d J = (\rd + G_3 \,\w)\d J =0~.}
The differential operator $\CD= \rd + G_3\, \w\,$ maps the two-form
$\d J$ into a direct sum of a three-form and a five-form.  More generally, it
maps the space of even forms to odd forms and vice verser.  These
types of operators have appeared in flux compactifications \GraMPT\
expressed in the pure spinor formalism of Hitchin \HitG.  The
three-form wedge $G_3\w $ is called a twist and we will call $\CD$ a
twisted differential operator.\foot{We note that $G_3$ is complex so the
twisted differential operator is also complex.  Also, we can consider 
replacing $\d J$ with a pure spinor of the form $\d(e^{iJ})$.} 
Now, if $\rd G_3 =0$, as is the case when the axion/dilaton $\t$ is a
constant, then it can be easily checked that $\CD\CD=0$.  We can
therefore treat the twisted differential operator as defining a generalized
cohomology.  Note that only a one-form can be mapped into a two-form
by the exterior derivative $\rd$ so $\d J$ would still be modded out
by an exact two-form.  Therefore, an interesting question is whether the
dimension of the generalized cohomology may be related to the
dimension of the moduli.

As for the heterotic case, constructing a generalized cohomology is
unfortunately not as straightforward.  The two constraints are 
\eqn\Hetm{\rd\, \d J =0 ~~,\qquad\quad -2{J_m}^n(\rd \,\d \p)_n = {1\o 2}
H_{mnp}{\d J}^{np}~.}
The constraints again can be combined together.  But first, we introduce the
following definitions.
\eqn\dcop{\rd^c = i(\bpa - \pa)~,}
\eqn\cont{H \llcorner J = {1 \o 2}(H_{mnp}\, g^{nr} g^{ps}J_{rs})\, \rd y^m~.}
The $\rd ^c$ operator is just $-{J_m}^n\pa_n$ in the
standard complex structure where ${J_a}^b=i{\d_a}^b$.  This operator
is natural in the heterotic theory since $\rd H \neq 0$ but we have
\eqn\Hdc{H=\rd^c J~~, \qquad\qquad \rd^c H= (\rd^c)^2 J =0~.}
Further, we note that the H-contraction operator $H \llcorner$ can also
be defined more generally to map a $p$-form into a $(p-1)$-form (for
$p\geq 2$).  The two
constraints in \Hetm\ can now be re-written as
\eqn\Hetma{\rd^c\, \d J=0 ~~, \qquad\quad 2\,\rd^c (\d \p) = - H
\llcorner \d J~,}
where the minus sign is due to $H_{mnp}\d J^{np}= -
H_{mnp}\,g^{nr}g^{ps}\d J_{rs}$.  Notice that the integrability condition of
the second constraint equation given in \vhetc\ can be simply expressed
as $\rd^c (H \llcorner \d J) =0 $.  We now formally combine the two
constraints as follows.
\eqn\hetmb{\CD \,(e^{2\d\p} e^{\d J})  = (\rd^c + H \llcorner) (e^{2\d \p}e^{\d J}) = 0~,}
where for $e^{2\d\p} e^{\d J}$, only the linear order in variation
should be kept, i.e. $e^{2\d\p} e^{\d J}=1+2\,\d\p + \d J$.   Here
again, the $\CD$ operator map even forms to odd forms and vice verser,
but it is no longer the case that $\CD\CD=0$.  Nevertheless, it is interesting
to point out that if we had in the IIB theory wrote the primitivity constraint
as $G_3\llcorner \d J=0$, then the IIB and heterotic twisted differential
operators would be similar except for exchanging $\rd \leftrightarrow
\rd^c$ and $G_3 \leftrightarrow H_3$.

\bigskip\bigskip\bigskip
\centerline{\bf Acknowledgements}
\medskip
It is a pleasure to thank R. Donagi, M. Douglas, S. Giddings, G. Moore,
D. Robbins, M. Schulz, S. Sethi, R. Tatar, W. Taylor, D. Toledo,
A. Tomasiello, and Y.-S. Wu.  We are particularly indebted to
M. Becker, K. Dasgupta, and E. Sharpe for numerous discussions on
heterotic flux models and to D. Waldram for explaining his works on
intrinsic torsion and pure spinors to us.  We also like to thank the
Aspen Center for Physics for hospitality during the latter stages of
this work.  This work is supported in part by NSF grant PHYS-0244722
and the University of Utah.

\listrefs
\end